\documentclass[10pt,aps,prl,preprintnumbers,superscriptaddress,showpacs,
twocolumn,floatfix,nofootinbib]{revtex4-2}
\usepackage{graphicx}
\usepackage{amsmath}
\usepackage{slashed}

\newcommand{\bfm}[1]{\mbox{\boldmath$#1$}}

\newcommand{\gsim}{\;\rlap{\lower 3.5 pt \hbox{$\mathchar \sim$}} \raise 1pt
\hbox {$>$}\;}
\newcommand{\lsim}{\;\rlap{\lower 3.5 pt \hbox{$\mathchar \sim$}} \raise 1pt
\hbox {$<$}\;}

\begin{document}

\title{\boldmath
Dynamical Confinement and Magnetic Traps for Charges and Spins
\unboldmath}
\author{Afshin Besharat}
\email[]{abeshara@ualberta.ca}
\affiliation{Department of Physics, University of Alberta, Edmonton, Alberta T6G
2J1, Canada}
\author{Alexander A. Penin}
\email[]{penin@ualberta.ca}
\affiliation{Department of Physics, University of Alberta, Edmonton, Alberta T6G
2J1, Canada}
%\affiliation{Institut f\"ur Theoretische Teilchenphysik,
%Karlsruher Institut f\"ur Technologie (KIT), 76128 Karlsruhe, Germany}
%\affiliation{Institute for Theoretical Physics, ETH Z\"urich, 8093 Z\"urich,
%Switzerland}
\begin{abstract}
We describe  a novel mechanism of  charged particles
confinement by a rapidly oscillating magnetic field. It relies
on the renowned dynamical stabilization phenomenon and provides
a foundation for a new class of the magnetic traps. The
dynamical magnetic confinement of charges and spin magnetic
moments has a number of remarkable properties which make it a
promising alternative  to the existing techniques in a wide
range of physical problems.
\end{abstract}
%\pacs{}
\preprint{ALBERTA-THY-1-25}

\maketitle

Electromagnetic traps  designed to operate  individual
particles and atoms play a key role in the solution of the
physical problems ranging from the measurement of the neutron
lifetime \cite{Huffman:2000fh} and the fine structure constant
\cite{Hanneke:2008tm} to synthesis  of antimatter
\cite{Andresen:2010} and quantum computing \cite{Monroe:2021}.
Depending on the problem they can rely on quite  diverse
physical principles
\cite{Penning:36,Kapitza:1951,Vladimirski:1960}. For example,
the dynamical stabilization \cite{Kapitza:1951} is used  in the
design of the Paul traps \cite{Paul:1990}, where the rapidly
oscillating electric field creates confining potential for
charged particles.  On the other hand,  the magnetic traps
\cite{Kugler:1978ae,Pritchard:1983} capture the neutral
particles through  the magnetic moment interaction   to the
static  spatially inhomogeneous magnetic field. Dynamical
confinement of magnetic moments has also been discussed in the
past \cite{Lovelace:1985,Lovelace:1987} and verified
experimentally \cite{Cornell:1991}.

The problem of localizing particles by electromagnetic field
has a long history and despite many significant advances still
poses a serious theoretical challenge as the available
solutions and techniques  have  a number of  fundamental
limitations.  For example, in the Penning traps
\cite{Brown:1985rh} the homogeneous magnetic  field  confines
the  radial motion of charged particles  near the {maximum} of
the potential energy in the transversal plane so that the
energy dissipation results in the particle loss. A similar
problem exists in the static magnetic traps for the neutral
particles with an intrinsic magnetic moment, where the
particles are not trapped at the total energy minimum and the
spin flip results in  the particles running away
\cite{Bergeman:1987}. At the same time the fine stability
conditions characteristic to the techniques relying on the
dynamical stabilization by the oscillating electric field
\cite{Paul:1990} prevents trapping distinct particle species.
Many outstanding technical solutions have been found to
circumvent these limitations \cite{Walz:1995,Petrich:1995,
Davis:1995,Fortagh:2007} leading to the breakthrough results in
different fields  \cite{Eades:1999,Leggett:2001}.  However the
core physical principles of the particle  traps remain the same
for a few decades and further progress may require  new ideas.
In this paper we describe a confinement mechanism based on the
dynamical stabilization caused by the fast oscillations of the
magnetic field suggesting a physical principle for a new class
of particle traps. The mechanism applies both to the electric
charges and spin magnetic moment, does not have the above
limitations,  and has a number of remarkable properties
relevant  for  the keynote applications including the study of
anti-hydrogen,  cold atoms, {\it etc}. We present the examples
of the confining potential for  charges and spin magnetic
moments which   can be realized experimentally with the
existing technology and setup. While the dynamical magnetic
confinement of electric charges has not been discussed so far,
we revise the previous analysis
\cite{Lovelace:1985,Lovelace:1987} of the spin magnetic
moments.

The theoretical description  of the new dynamical stabilization
phenomenon does not reduce to the standard Floquet/Mathieu
analysis. It requires a new approach applicable to the
intrinsically  three-dimensional systems with the interaction
depending on particle velocity.  Despite  long-time efforts the
relevant framework has been formulated only recently
\cite{Penin:2023dtn} inspired by the ideas of the effective
field theory  in  particle physics. In this Letter we extend
this method to the general case of  nonconservative and
velocity-dependent forces.

We start with the general description of classical and quantum
dynamics  in the  rapidly oscillating  magnetic field. The
theory of the periodically driven systems in the high-frequency
limit is based on the concept of averaging, when the effect of
the oscillating field is smeared out and the long-time
evolution is governed by the resulting effective interaction.
The method is well known in classical mechanics
\cite{Bogoliubov:1961}. It has been extended to quantum systems
\cite{Cook:1985,Grozdanov:1988,Rahav:2003a,Rahav:2003b} and
refined and generalized in many subsequent works
\cite{Verdeny:2013,Goldman:2014,Eckardt:2015,Itin:2015,
Mikami:2016,Bukov:2016,Weinberg:2017,Restrepo:2017}.
Here we adopt the effective field theory
approach developed in \cite{Penin:2023dtn} to systematically
derive  the effective action to any order of the {\it
high-frequency expansion} in the ratio of the oscillation
period to a characteristic time scale of the averaged system.
Let us discuss first the classical system of a particle of mass
$m$ and electric charge $e$ subject to the  Lorentz force
\begin{equation}
\begin{split}
&{\bfm F}(t,{\bfm r})=e\left({\bfm E}(t,{\bfm r})
+{\bfm v}\times {\bfm B}(t,{\bfm r})\right)
\label{eq::force}
\end{split}
\end{equation}
due to the oscillating magnetic field  ${\bfm B}(t,{\bfm
r})=\cos(\omega t){\bfm B}({\bfm r})$ and electric field ${\bfm
E}(t,{\bfm r})=\sin(\omega t){\bfm E}({\bfm r})$, where the
bold fonts indicate  three-dimensional vectors.  For the
magnetic field generated by an external source, in the region
of vanishing  charge and current density the Maxwell equations
impose the relations ${\bfm \partial}\times{\bfm E}({\bfm
r})=\omega{\bfm B}({\bfm r})$, ${\bfm \partial}\times{\bfm
B}({\bfm r})={\cal O} (1/c^2)$, where $c$ is the speed of
light. We consider the case when the electromagnetic radiation,
{\it i.e.} the  curl of the magnetic field,  can be neglected.
Following \cite{Penin:2023dtn} we split  the particle
coordinates  into the slow and fast modes
\begin{equation}
\begin{split}
{\bfm r}\to {\bfm r}+\sum_{n=1}^\infty
\left[\bfm c_n({\bfm r})\cos(n\omega t)
+\bfm s_n({\bfm r})\sin(n\omega t)\right]\,,
\label{eq::modes}
\end{split}
\end{equation}
where the vector ${\bfm r}$ now describes the  slow smeared
motion, and split the total time derivative  into the slow and
fast components ${d/dt}={\bfm
v}\cdot\bfm\partial_r+\partial_t$. Substituting this
decomposition  into the equation of motion and reexpanding in
the Fourier harmonics one can find the coefficients $\bfm
c_n({\bfm r})$ and $\bfm s_n({\bfm r})$ order by order in
$1/\omega^2$. The zero harmonic then defines the equation of
motion for the slow  ``time-averaged''  evolution. The
corresponding  effective Lagrangian through the next-to-leading
order of the high-frequency  expansion reads
\cite{Penin:2023dtn}
\begin{equation}
\begin{split}
&{\cal L}_{\rm eff}={m\over 2}{v_iv_j}g_{ij}({\bfm r})
-V_{\rm eff}({\bfm r})+{\cal O}(1/\omega^6),
\label{eq::Leff}
\end{split}
\end{equation}
where ${\bfm v}={d{\bfm r}/dt}$, and the summation over the
repeating indices is implied. Here $g_{ij}$ is the induced
three-dimensional metric and $V_{\rm eff}$ is  the effective
potential. Keeping the leading   terms  quadratic in the
electric and  magnetic fields and eliminating the latter by the
relation ${\bfm B} = {\bfm \partial}\times{\bfm E}/\omega$ we
get
\begin{eqnarray}
g_{ij}&=&\delta_{ij} -
{e^2\over 2m^2\omega^4}\left({\bfm \partial}E_i{\bfm \partial}E_j
+\partial_i{\bfm E}{\bfm \partial}E_j
+\partial_j{\bfm E}{\bfm \partial}E_i\right)
\nonumber\\[1mm]
&&+{\cal O}(1/\omega^6)
\label{eq::metric}
\end{eqnarray}
and
\begin{equation}
\begin{split}
&V_{\rm eff}={e^2\over 4m\omega^2}{\bfm E}^2+{\cal O}(1/\omega^6),
\label{eq::Veffclass}
\end{split}
\end{equation}
where ${\bfm B}$ and ${\bfm E}$ stand for the field amplitudes
${\bfm B}({\bfm r})$ and ${\bfm E}({\bfm r})$, respectively.
The leading order effective potential has the same form as in
the case of the oscillating electric field when the  magnetic
field can be neglected. However, its  properties are quite
different since the magnetically induced electric field is not
potential. Moreover, for a fixed value of ${\bfm B}$ the
magnitude  of the induced electric field grows linearly with
$\omega$ and the effective potential remains finite at
$\omega\to \infty$. At the same time the induced metric
Eq.~(\ref{eq::metric}) reduces to the result for a potential
field \cite{Penin:2023dtn} only for ${\bfm B}=0$.

Let us now consider the corresponding quantum system with the
time-dependent  Hamiltonian
\begin{equation}
\begin{split}
{\cal H}(t)={1\over 2m}\left(\hat{\bfm p}-e{\bfm A}(t,{\bfm
r})\right)^2+eV(t,{\bfm r})-\hat{\bfm \mu}{\bfm B}(t,{\bfm r}),
\label{eq::Ham}
\end{split}
\end{equation}
where $\hat{\bfm p}=-i\hbar\bfm\partial$, $\hat{\bfm
\mu}={g\mu\over \hbar}\hat{\bfm S}$ is the  magnetic moment,
$g$ is the gyromagnetic ratio, $\mu$ is the spin magnetic
moment unit (magneton), and $\hat{\bfm S}$ is the operator of
the spin. The vector and scalar potentials are given by ${\bfm
A}(t,{\bfm r})=\cos(\omega t){\bfm A}({\bfm r})$, $V(t,{\bfm
r})=\sin(\omega t)V({\bfm r})$ so that the electric field
amplitude is  ${\bfm E}({\bfm r})=\omega {\bfm A}({\bfm
r})-{\bfm \partial}V({\bfm r})$. The effective time-independent
Hamiltonian describing the low-energy excitations of the system
can be found by the expansion of the Schr\"odinger equation
Green's function ${\cal G}=(i\hbar\partial_t-{\cal
H}(t)+i\varepsilon)^{-1}$ in the inverse powers of $\omega^2$
in Fourier space, similar to the nonrelativistic expansion of
the massive  Dirac propagator in an external field.  For the
spin-independent potential interaction the resulting Feynman
rules of the {\it high-frequency effective theory} (HFET)  have
been derived through higher orders of the  expansion
\cite{Penin:2023dtn}. The generalization of the analysis to the
magnetic and spin interaction is rather straightforward. Let us
outline the calculation of the spin-dependent part. The Fourier
transform  $\tilde {\cal G}({\cal E};\bfm p_i,\bfm p_f)$ of the
Green's function depends on the energy, initial and final
momenta satisfying the high-frequency condition ${\cal E},~\bfm
p_{i,f}^2/m\ll \hbar \omega$. Expanding it in powers of the
magnetic moment we get
\begin{equation}
\begin{split}
&\tilde {\cal G}=\tilde {\cal G}_0
+\tilde {\cal G}_0 \,\, \hat{\bfm \mu}\tilde{\bfm B} \,\,
\tilde {\cal G}_0\,\,\hat{\bfm \mu}\tilde{\bfm B}\,\,
\tilde {\cal G}_0+\ldots\,,
\label{eq::GFseries}
\end{split}
\end{equation}
where $\tilde {\cal G}_0({\cal E},\bfm p)=({\cal E}-\bfm
p^2/2+i\varepsilon)^{-1}$ is the free particle propagator,
$\tilde{\bfm B}$   is the Fourier transform of ${\bfm
B}(t,{\bfm r})$, and the term linear in the external field
vanishes by energy conservation due to the condition $E\ll
\hbar\omega$. In the second term of Eq.~(\ref{eq::GFseries})
the intermediate state propagator carrying a momentum ${\bfm
p}$ and the energy ${\cal E}+\hbar\omega$ is far off-shell and
can be expanded  as follows
\begin{equation}
\begin{split}
&\tilde {\cal G}_0({\cal E}+\hbar\omega, {\bfm p})
={1\over \hbar\omega}-{{\cal E}
- \bfm p^2/(2m)\over (\hbar\omega)^2}+\ldots\,,
\label{eq::GF0series}
\end{split}
\end{equation}
giving  rise to a local effective interaction, quadratic in the
external field. The contribution of the first term in
Eq.~(\ref{eq::GF0series}) as well as all odd negative powers of
$\omega$ vanish due to the time-reversal symmetry, while the
second term results in a spin-dependent  {\it seagull} HFET
vertex. Its matrix element between the free on-shell states
reads
\begin{equation}
\begin{split}
&{1\over 2(\hbar\omega)^2}
\langle {\cal E},{\bfm p}_f|(\hat{\bfm \mu}{\bfm B})
({\hat{\bfm p}^2/(2m)}-{\cal E})
(\hat{\bfm \mu}{\bfm B})|{\cal E},{\bfm p}_i\rangle
\\&={1\over 4m\omega^2}
\langle {\cal E},{\bfm p}_f|\left(\hat{\bfm \mu}
\partial_i{\bfm B}\right)^2|{\cal E},{\bfm p}_i\rangle\,,
\label{eq::vertexlo}
\end{split}
\end{equation}
where we used the on-shell condition ${\bfm p}_{i,f}^2=2m{\cal
E}$. This vertex corresponds to the ${1\over
4m\omega^2}\left(\hat{\bfm \mu}\partial_i{\bfm B}\right)^2$
spin-dependent term in the effective  Hamiltonian. The
contribution of the  vector and scalar potential coupled to the
particle charge in Eq.~(\ref{eq::Ham}) can be obtained in the
same way, with the full HFET Hamiltonian given by
\begin{equation}
\begin{split}
&{\cal H}_{\rm eff}={1\over 2m}
\hat{p}_ig^{-1}_{ij}({\bfm r})\hat{p}_j
+V_{\rm eff}({\bfm r})+{\cal O}(1/\omega^6),
\label{eq::Heff}
\end{split}
\end{equation}
where $g^{-1}_{ij}$ is the inverse of the metric tensor
Eq.~(\ref{eq::metric}), the gauge-invariant effective
potential includes  the quantum corrections
\begin{eqnarray}
V_{\rm eff}&=&{1\over 4m\omega^2}\left[e^2{\bfm E}^2
+\left(\hat{\bfm \mu}\partial_i{\bfm B}\right)^2\right]
\nonumber\\
&-&{\hbar^2e^2\over 8m^3\omega^4}({\bfm \partial}\partial_j{E}_i)
({\bfm \partial}\partial_i{E}_j)+{\cal O}(1/\omega^6),
\label{eq::Veff}
\end{eqnarray}
and the gauge-invariant interaction to a time-independent or
slowly varying electromagnetic potential can be included in the
standard way. In the  limit $\hbar\to 0$, Eq.~(\ref{eq::Heff})
reproduces the classical result Eq.~(\ref{eq::Leff}). For the
degenerate $S=1/2$ states the second term of
Eq.~(\ref{eq::Veff}) becomes ${1\over 4m\omega^2} \left({g
\mu\over 2}\right)^2 \left(\partial_i\bfm B\right)^2$. If the
magnetic moment has a definite projection on a given axis, the
vector ${\bfm B}$ in this expression should be replaced by its
projection on the magnetic moment direction. This occurs, {\it
e.g.} when a static homogeneous magnetic field ${\bfm B}_0$ is
applied to the system,  with the Larmor frequency exceeding the
driving frequency $\omega$. This case has been discussed in
\cite{Lovelace:1985,Lovelace:1987}, where the authors adopted
$-\mu |{\bfm B}_0+{\bfm B}(t,{\bfm r})|$ as the time-dependent
interaction in Eq.~(\ref{eq::Ham}). This approximation,
however, is not consistent since the time averaging applies to
the individual components  of the oscillating magnetic field
rather than to its absolute value. As a result, the effective
potential derived in \cite{Lovelace:1985,Lovelace:1987} agrees
with Eq.~(\ref{eq::Veff}) only in the limit $|{\bfm B}(t,{\bfm
r})|/|{\bfm B}_0|\to 0$. In the existing experimental setup
\cite{Cornell:1991} this condition is satisfied only near the
minimum of the potential, and the result
\cite{Lovelace:1985,Lovelace:1987} cannot be used {\it e.g.} to
evaluate the actual shape and depth of the confining potential.

\begin{widetext}

\begin{figure}[t]
\begin{center}
\begin{tabular}{ccc}
\hspace*{-3mm}\includegraphics[width=6.0cm]{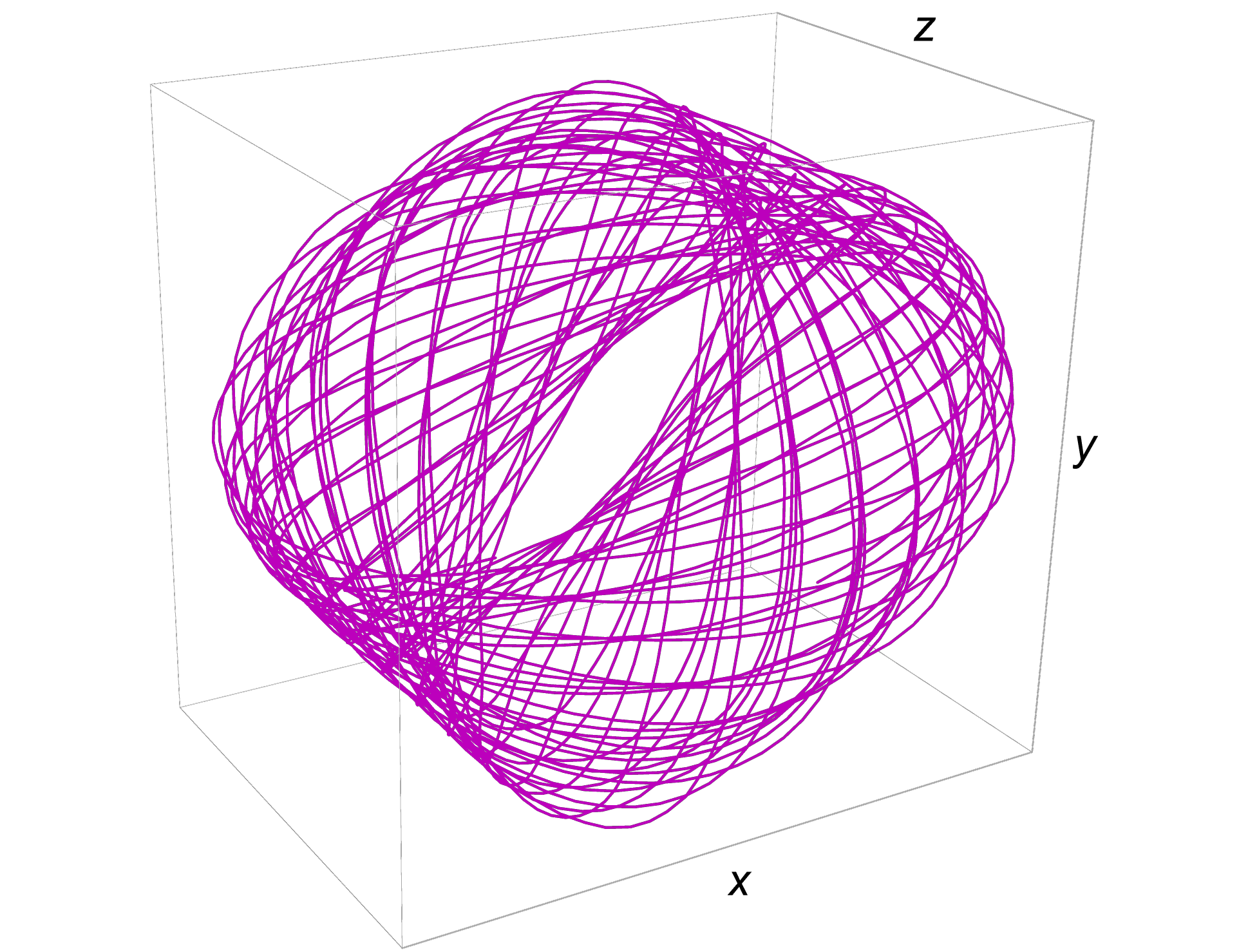}&
\hspace*{0mm}\includegraphics[width=6.0cm]{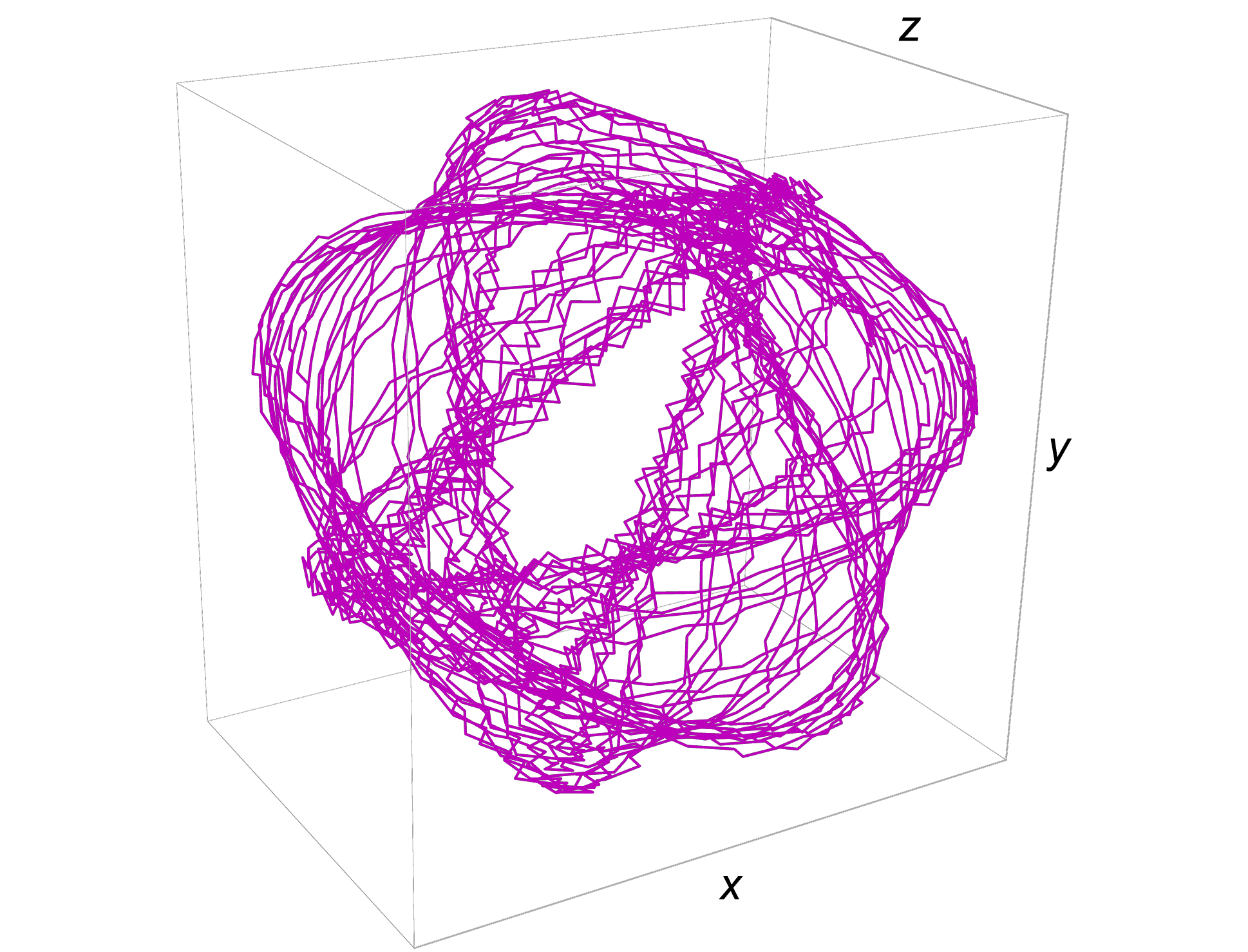}&
\hspace*{0mm}\includegraphics[width=6.0cm]{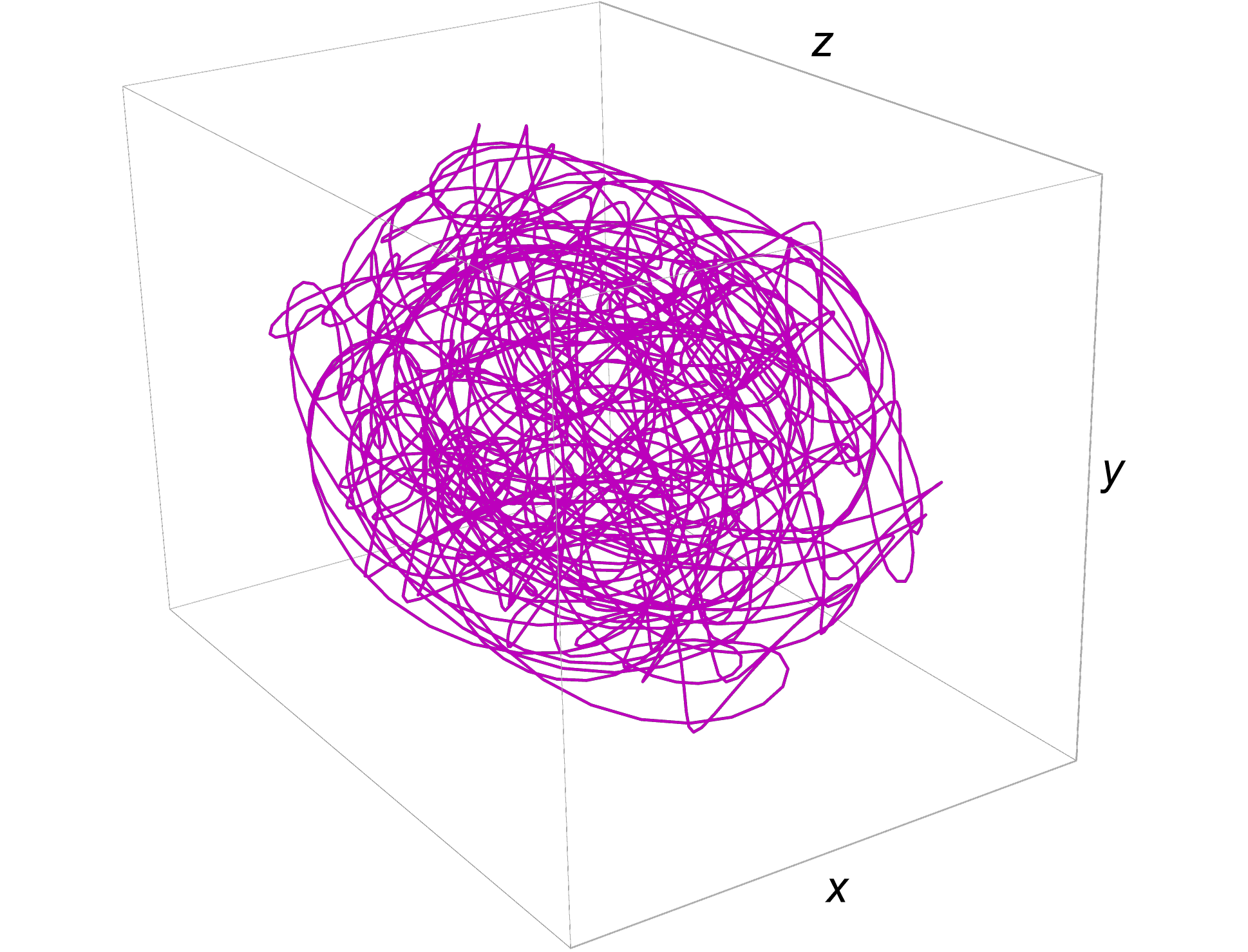}\\[5mm]
\hspace*{-3mm}(a)&
\hspace*{0mm}(b)&
\hspace*{0mm}(c)\\
\end{tabular}
\end{center}
\caption{\label{fig::1} The   trajectories of a charged
particle in the three-dimensional magnetic trap for (a)
$\omega_B/\omega =1/100$,  (b) $\omega_B/\omega =1/10$, and  (c)
$\omega_B/\omega =1$. As the scale ratio increases, the motion
transforms from quasiperiodic to chaotic, but remains confined.}
\end{figure}

\end{widetext}

There exist various shapes of the  spatial distribution of the
oscillating field amplitude, which make the effective potential
Eq.~(\ref{eq::Veff}) confining  for charges and spins. The
simplest realization of the  charge confinement in two
dimensions is given  by a locally homogeneous amplitude ${\bfm
B}= \left(0,0,B\right)$, where we assume an axially symmetric
region of a finite cross section with the flux of the uniformly
oscillating magnetic field rapidly vanishing  at the
boundaries. This is exactly the field inside a sufficiently
long solenoid, which can easily by realized in practice. Then
the induced electric field vanishes on the $z$-axis, and away
from the boundaries is given by  ${\bfm E}={\omega B\over 2}
\left(-y,x,0\right)$. In the high-frequency limit the leading
effect of the oscillating field reduces to the  first term in
Eq.~(\ref{eq::Veff}), which gives
\begin{equation}
\begin{split}
&V_{\rm eff}={m\omega_B^2\over 16}\left(x^2+y^2\right),
\label{eq::Veffcharge}
\end{split}
\end{equation}
where $\omega_{B}=eB/m$ is the cyclotron frequency associated
with the field oscillation amplitude. Note that the resulting
two-dimensional harmonic oscillator has the frequency
$\omega_B/2^{3/2}$ rather than $\omega_{B}$ appearing in the
case of the static field. The effective potential
Eq.~(\ref{eq::Veffcharge}) does not depend explicitly on the
driving frequency $\omega$ but the convergence of the
high-frequency expansion formally requires $\omega \gg
\omega_B$,  which may set a practical limit on the magnitude of
the oscillating field and, therefore, on the binding strength.
At the same time the calculation of the HFET action through
${\cal O}(1/\omega^6)$ \cite{Penin:2023dtn} indicates that the
high-frequency expansion is not plagued by the large numerical
coefficients. For the above system its convergence  can be
estimated from Eq.~(\ref{eq::metric}), which reduces to  the
mass renormalization  factor $1+\omega_B^2/(8\omega^2)$ for the
motion in the transverse plane. Thus, the actual expansion
parameter  is $\omega_B^2/(8\omega^2)$, and the condition
$\omega\gsim \omega_B $ may be sufficient for the convergence
and the stability of the confining potential as in the case of
one-dimensional  Mathieu equation.\footnote{For the same set of
parameters the transition curve of Mathieu equation gives the
bound $\omega\ge 1.10\ldots \omega_B$.}

Though in the derivation of Eq.~(\ref{eq::Veffcharge}) we
assume a locally homogeneous axially symmetric magnetic field,
the existence of a minimum of the effective potential is
{topologically protected} against the spatial perturbations.
Indeed, the circulation of the induced electric field  implies
the existence of at least one ``vortex line'' where the
effective potential vanishes.

\begin{widetext}

\begin{figure}[t]
\begin{center}
\begin{tabular}{ccc}
\hspace*{-3mm}\includegraphics[width=5.0cm]{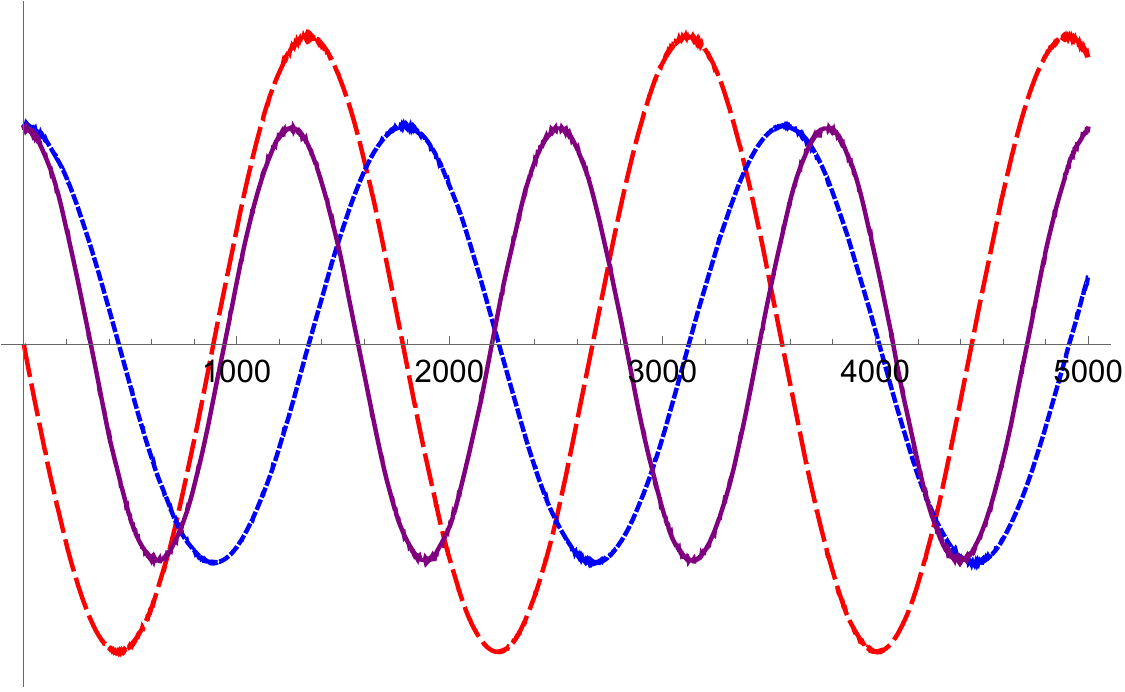}&
\hspace*{5mm}\includegraphics[width=5.0cm]{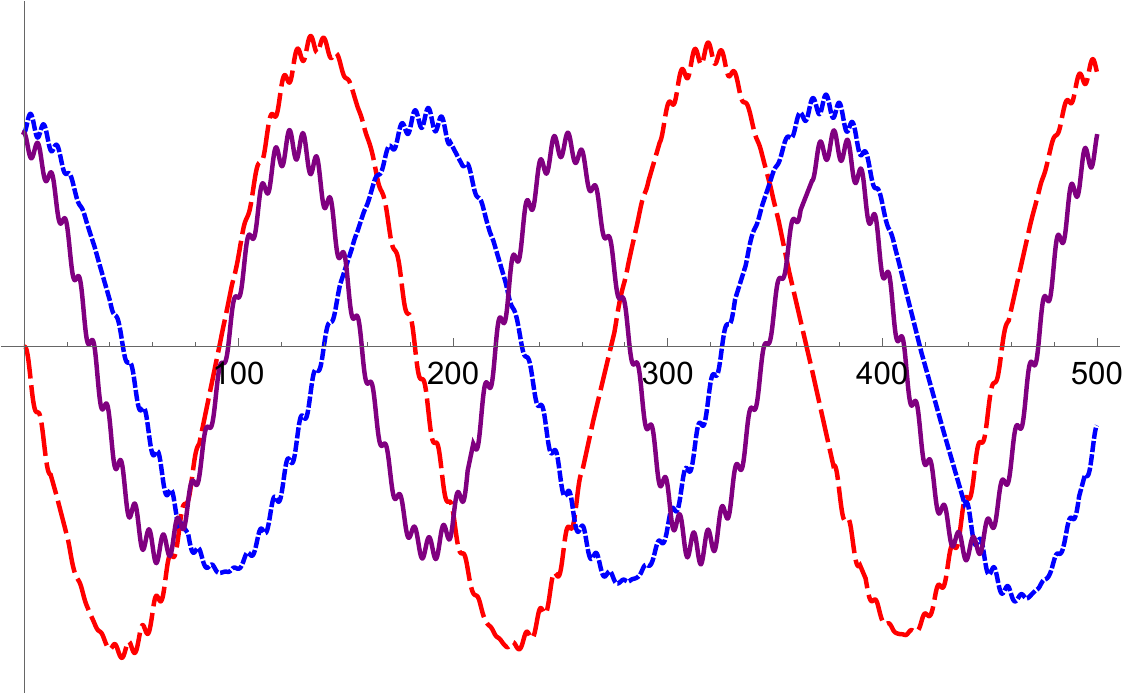}&
\hspace*{5mm}\includegraphics[width=5.0cm]{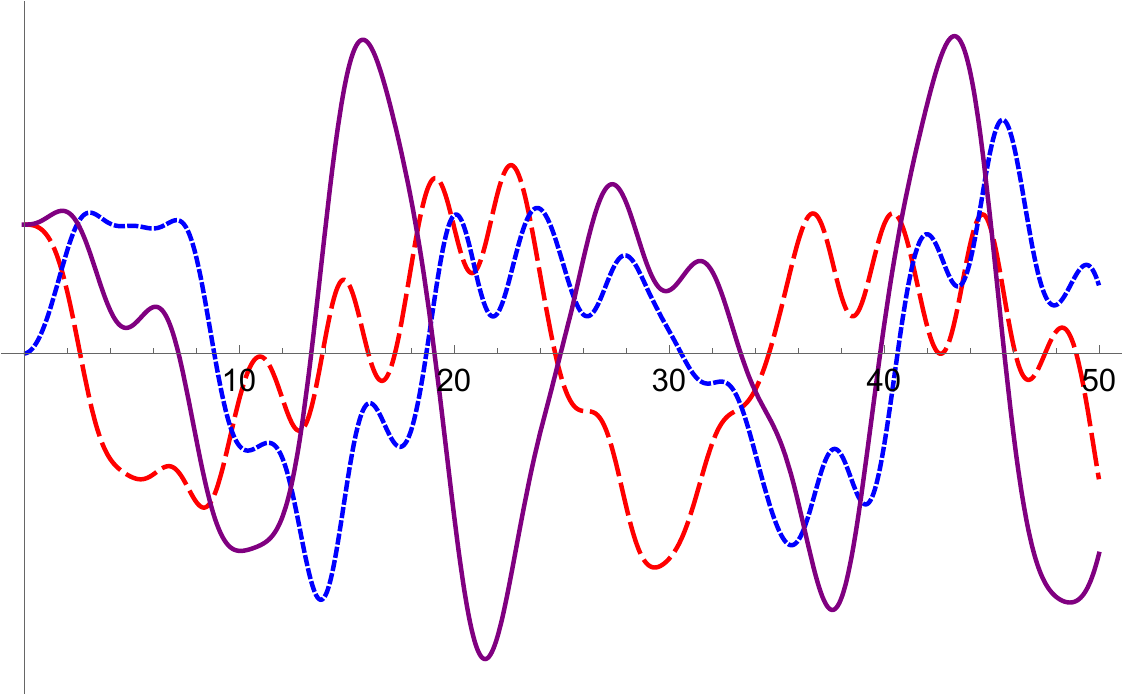}\\[5mm]
\hspace*{-3mm}(a)&
\hspace*{5mm}(b)&
\hspace*{5mm}(c)\\
\end{tabular}
\end{center}
\caption{\label{fig::2} The coordinates $x$ (short dashed), $y$
(long dashed), and $z$  (solid line)  of a charged particle in
the three-dimensional magnetic trap as functions of $\omega t$
for (a) $\omega_B/\omega=1/100$,  (b) $\omega_B/\omega =1/10$,
and  (c) $\omega_B/\omega =1$, corresponding to the trajectories in Fig.~\ref{fig::1}. For $\omega\gg\omega_B$ the large-scale oscillation frequencies are $\omega_B/2^{3/2}$
and  $\omega_B/2$, in agreement with Eq.~(\ref{eq::Veffcharge3d}).}
\end{figure}

\end{widetext}

\begin{figure}[t]
\begin{center}
\hspace*{0mm}\includegraphics[width=3.5cm]{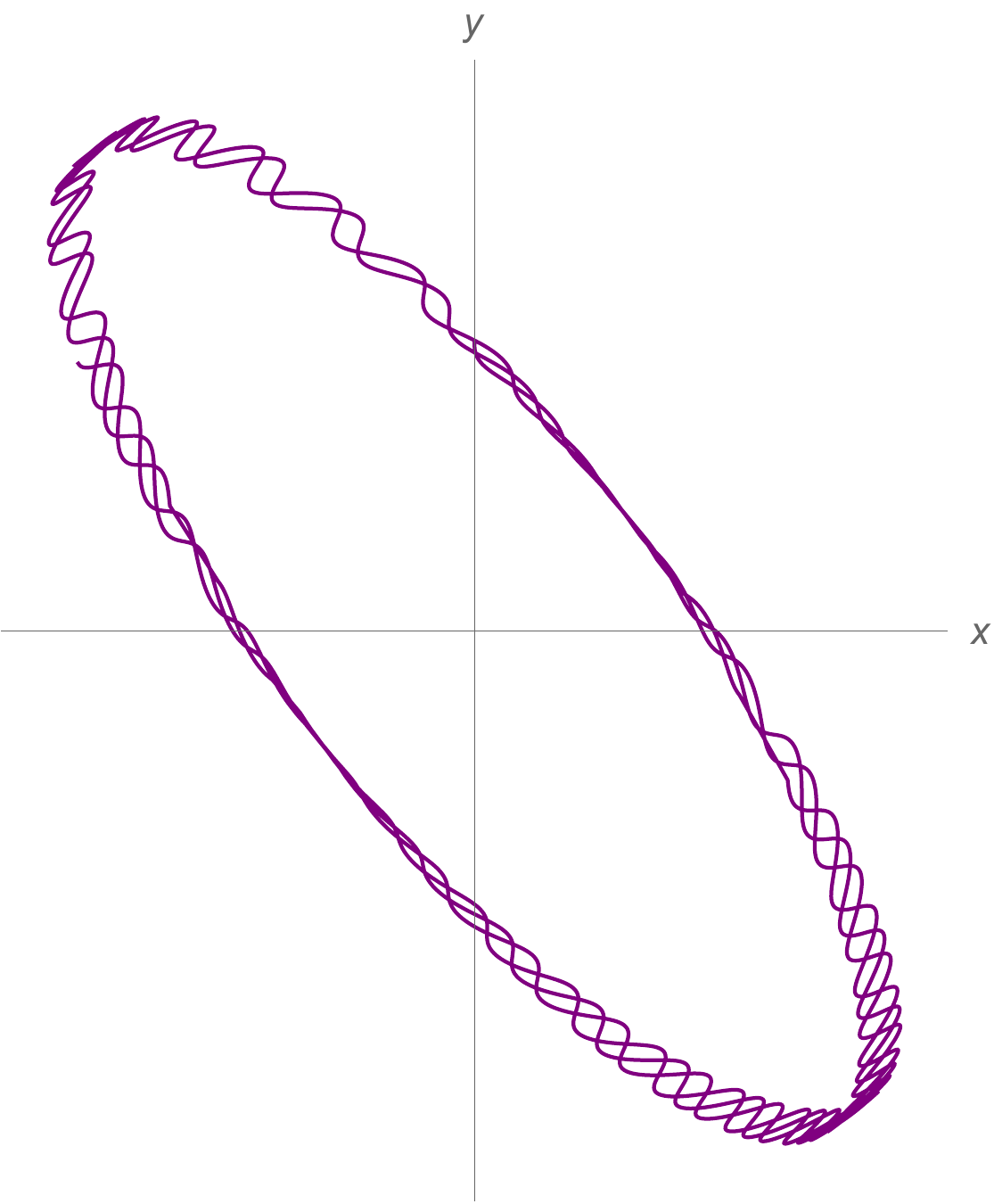}
\end{center}
\caption{\label{fig::3} A trajectory of a charged particle in
the two-dimensional magnetic trap for  $\omega_B/\omega =1/10$. The excess micromotion orthogonal to the displacement from the induced electric field nodal point at the origin can be clearly identified.}
\end{figure}

In three dimensions the confinement can be realized by a planar
rotating  field  ${\bfm B}(t,{\bfm r})= B\left(\cos(\omega
t),-\sin(\omega t),0\right)$ with the corresponding induced
electric field
\begin{equation}
\begin{split}
&{\bfm E}(t,{\bfm r})\!=\!{\omega B\over
2}\!\left(z\cos(\omega t), -z\sin(\omega t),y\sin(\omega
t)\!-\!x\cos(\omega t)\right)\!.
\label{eq::E3d}
\end{split}
\end{equation}
In principle the phase shift between the field components
requires a generalization of the  analysis given above.
However, to the leading order in $1/\omega^2$, the
generalization  is straightforward since the oscillation modes
do not interfere (their product averages to zero) and the
effective potential reads
\begin{equation}
\begin{split}
&V_{\rm eff}={m\omega_B^2\over 16}\left(x^2+y^2+2z^2\right).
\label{eq::Veffcharge3d}
\end{split}
\end{equation}
We have verified  our result with the numerical solution of the
exact time-dependent equations of motion. Characteristic shapes
of a particle trajectory in the magnetic trap for different
values of  the ratio $\omega_B/\omega$ are given in
Fig.~\ref{fig::1}. For $\omega_B/\omega\ll 1$ the motion
becomes quasiperiodic with the frequencies $\omega_B/2^{3/2}$
and  $\omega_B/2$,  in full agreement with
Eq.~(\ref{eq::Veffcharge3d}), see Fig.~\ref{fig::2}. As the
scale ratio increases  the motion becomes {\it chaotic} but
remains  {\it confined}  up to $\omega_B/\omega\approx 1$,
confirming our analysis of the high-frequency expansion
convergence. The numerical simulations also show that, in
contrast to the Paul traps, due  to the circulation of the
induced electric field the {excess micromotion}
\cite{Berkeland:1998} in the magnetic trap is orthogonal to the
displacement from the field nodal point, Fig.~\ref{fig::3}.

We can estimate  the binding energy of the trap by evaluating
the potential Eq.~(\ref{eq::Veffcharge3d}) at the scale of its
geometrical size $L$, which gives $E_{\rm bind}\approx
(eBL)^2/(16m)$. The depth of the corresponding potential well
is proportional to $\omega_B\Phi$, where $\Phi$ is the
amplitude of the total magnetic flux. For $L=1$~cm, the depth
of $1$~V is achieved with   $B\approx 400$~G, $\omega\approx
600$~kHz for a proton, and with $B\approx 10$~G, $\omega\approx
28$~MHz for an electron. In general, for two particle with the
same absolute value of electric charge but essentially
different masses $m_1\gg m_2$, the same binding requires
different magnetic fields   $B_1/B_2=(m_1/m_2)^{1/2}$ and
driving frequencies. Let us now consider a superposition of
such   modes in combination with a large static axial
homogeneous magnetic field $B_0\gg B_{1,2}$. This implies  the
hierarchy of the time scales
$\omega_B(B_0,m_2)\gg\omega_2\gsim\omega_B(B_2,m_2)\gg
\omega_1\gsim\omega_B(B_1,m_1)$. Then for the heavy particle
the effect of the fast precession of the resulting field with
the amplitude $B_2$  can be neglected. At the same time for the
light particle dynamics at the scale $\omega_2$,  the slow
precession  with the  amplitude  $B_1$ can be treated as a
small adiabatic variation of the background magnetic field.
Thus, we can perform the time averaging and get the confining
potential for the two modes independently. The presence of the
large background field  is mandatory. Indeed, for $B_0=0$ the
slowly oscillating  cyclotron frequency $\omega_B(B_1,m_2)$ at
some moment approaches the driving frequency $\omega_2$, which
may destabilize the motion of the light particle. For  $B_0\gg
B_1$ the cyclotron frequency corresponding to the slowly
precessing sum of $B_0$ and $B_1$ fields is of order
$\omega_B(B_0,m_2)$ at any time, which prevents the
stability loss. The above setup provides an analog of the
combined Paul-Penning   trap \cite{Walz:1995} or two-frequency
Paul trap \cite{Foot:2018} without static  or alternating
electric potentials, which  can be used to simultaneously trap
different particle species.

Remarkably, the rotating magnetic field has already been
engineered in the TOP traps for the magnetic moment of cold
atoms \cite{Petrich:1995}. With the given values of the
parameters and with unequal oscillation frequencies of
the  magnetic field components  it would be able to actually
trap ions, though due to a relatively weak magnetic field the
corresponding potential well is rather shallow, {\it e.g.} for
a proton it  is only about $10^{-3}$~V.

The confining mechanism for the neutral particles is quite
different and is determined by the second  term in
Eq.~(\ref{eq::Veff}). Hence, it requires a spatially
inhomogeneous magnetic field amplitude.  As an example, let us
consider   an axially symmetric magnetic field of the commonly
used Ioffe-Pritchard traps \cite{Pritchard:1983}
\begin{equation}
\begin{split}
&{\bfm B}=B\left({xz\over\Delta^2}, {yz\over\Delta^2},
1+{x^2+y^2-2z^2\over 2\Delta^2}\right),
\label{eq::Bfield}
\end{split}
\end{equation}
where  $B$ is the value of the homogeneous component of the
field and an adjustable parameter $\Delta$ defines  the scale
of the field variation determined by the trap geometry.
%For the oscillating field  with the amplitude given by
%Eq.~(\ref{eq::Bfield}) the induced electric field amplitude
%reads
%\begin{equation}
%\begin{split}
%&{\bfm E}={\omega B\over 2}
%\left(y\left(1+{x^2-z^2\over\Delta^2}\right),
%x\left(-1+{z^2-y^2\over\Delta^2}\right),0\right).
%\label{eq::Efield}
%\end{split}
%\end{equation}
The corresponding effective potential  is
\begin{equation}
\begin{split}
&V_{\rm eff}={m\tilde{\omega}^2\over 2}\left(x^2+y^2+3z^2\right),
\label{eq::Veffspin}
\end{split}
\end{equation}
where
\begin{equation}
\begin{split}
& \tilde{\omega}={g\mu B\over  2m \Delta^2 \omega}.
\label{eq::omegatilde}
\end{split}
\end{equation}
Thus, the magnetic moment is harmonically trapped at the
origin. Eq.~(\ref{eq::Veffspin}) does not depend on the
homogeneous component of the field and is a function of the
field curvature $B/\Delta^2$ only.  If a static homogeneous
magnetic field is applied to the system, with the Larmor
frequency exceeding $\omega$,  the effective potential has to
be adjusted since only the projection of Eq.~(\ref{eq::Bfield})
on the static field direction contributes. For example,  with
the static field applied in the axial direction, the expression
in the brackets in Eq.~(\ref{eq::Veffspin}) should be replaced
by ${x^2/2}+{y^2/2}+2z^2$. In this case the equations of motion
following from the Hamiltonian Eq.~(\ref{eq::Ham})   reduce to
the linear Mathieu equations and the stability of the particle
equilibrium at the minimum of the effective potential can be
analysed beyond the high-frequency expansion. This gives a low
bound on the driving frequency $\omega_{\rm min}=\left({g\mu
B\over  \kappa m \Delta^2}\right)^{1/2}$, where
$\kappa=0.454\ldots$ is the root of the Mathieu equation
transition curve \cite{Kovacic:2018}. Numerically, for
$B/\Delta^2=10^3~{\rm G/cm}^2$ we get $\omega_{\rm min}
/(2\pi)\approx 790$~Hz for a  hydrogen  atom and about $160$~Hz
for sodium, which may be well within the experimental reach
\cite{Cornell:1991}. The corresponding binding energy can be
estimated  by evaluating the effective potential
Eq.~(\ref{eq::Veffspin}) at ${\bfm r}=(\Delta,0,0)$ and
$\omega=\omega_{\rm min}$ with the result    $E_{\rm bind}
\approx 0.1\, \mu B$. Thus, the binding energy which  can be
achieved by dynamical stabilization is parameterically  the
same as for the static magnetic traps with similar integral
spatial variation of the magnetic field, though the stability
constraint results in a numerical suppression factor which
weakly depends on the trap geometry. The main advantage of the
dynamical confinement, however, is that it  does not depend on
the particle spin orientation and traps the spins at the
absolute energy minimum preventing the loss of the particles
due to the spin flip \cite{Pritchard:1983,Lovelace:1985,
Lovelace:1987,Cornell:1991}.

To summarize, we have described a new   mechanism of charged
particles confinement through the dynamical stabilization  by
rapidly oscillating magnetic field, suggesting  a physical
principle for a new class of the particle traps. It   does not
involve a static or alternating electric potential and is thus
free of many intrinsic limitations of the existing types of the
traps.  The  dynamical magnetic traps possess some distinct
features briefly outlined below, which make them a promising
alternative to the available techniques in a number of the  key
physical applications.  They can be designed to confine
different  particle species simultaneously in the same region
of space. The purely harmonic confining potential for electric
charges  is generated  by an oscillating (for two dimensions)
or rotating (for three dimensions) locally homogeneous magnetic
field, similar to the existing TOP design \cite{Petrich:1995}.
The existence of its local minimum  is {\it topologically
protected} with respect to the spatial field perturbations
and does not require a  precise trap geometry.  Its stability is
entirely controlled by the driving frequency $\omega$, which
has to exceed the cyclotron frequency $\omega_B$ associated
with the amplitude of the oscillating field. The depth of the
trap is determined by the product of $\omega_B$ and  the
magnetic flux amplitude. Unlike the Paul traps \cite{Paul:1990}
in the stability region it  does not depend on $\omega$ and can
in principle  be  {\it scaled up arbitrarily}  with the total
magnetic flux.

Finally, we would like to note that the theoretical analysis of
the dynamical magnetic  confinement required a nontrivial
extension  of the recently developed effective field theory
description of the classical and quantum systems embedded in
the rapidly oscillation fields \cite{Penin:2023dtn} to the
general case of the nonpotential interactions. Within this approach
we were able to revise the long-established analysis of the
dynamical confinement of the a spin magnetic moments and
derived the first correct and general expression for the
corresponding confining   potential as well as the upper bound
on the magnetic trap depth.

\vspace{5mm}
\noindent
{\bf Acknowledgments.} The work of A.B. is supported by NSERC.
The work of A.P. was supported in part by NSERC and the
Perimeter Institute for Theoretical Physics.  A.P.
is thankful to the Munich Institute for Astro-, Particle and
BioPhysics (MIAPbP), funded by the DFG under Germany’s
Excellence Strategy – EXC-2094-390783311, where part of this
work was completed.

%%%%%%%%%%%%%%%%%%%%%%%%%%%%%%%%%%%%%%%%%%%%%%%%%%%%%%%%%%%%

\end{document}